\documentclass[11pt]{article}

\usepackage{dsfont}
\usepackage{subcaption}
\usepackage{fullpage}
\usepackage{graphicx}
\usepackage{mathrsfs}
\usepackage{amsthm}
\usepackage{amsmath}
\usepackage{amssymb}
\usepackage{booktabs}
\usepackage[usenames, dvipsnames]{color}
\theoremstyle{definition}

\newtheorem{definition}{Definition}[section]

\usepackage[round]{natbib}
\begin{document}
\title{Monte Carlo Simulations on robustness of functional location estimator based on several functional depth}
\author{Xudong Zhang}
\maketitle

\begin{abstract} 
Functional data analysis  has been a growing field of study in recent decades, and one fundamental task in functional data analysis is estimating the sample location.  
A notion called statistical depth has been extended from multivariate data to functional data, and it can provide a center-outward order for each observation within a sample of functional curves.
Making use of this intuitive nature of depth methods, a depth-based trimmed mean where curves with lower depth values are excluded can be used as a robust location estimator for the sample.
In this project, we first introduced several state-of-the-art depth approaches for functional data. These depths were half region depth, functional majority depth, band  depth, modified band depth and functional spatial depth. Then we described a robust location estimator based on functional depth, and studied performances of these estimators based on different functional depth approaches via simulation tests. Finally, the test results showed that estimators based on functional spatial depth and modified band depth exhibited superior performances.
\end{abstract}

\section{Introduction}
\paragraph{ } 
Functional data emerged frequently in a variety of fields such as economics, biology, meteorology, etc. It has been a growing field of study in recent decades. Each functional observation represents a functional curve over a continuum such as time, length or volume. One of the most obvious and perhaps the most important features is that the dimension of functional data is theoretically infinite. Although in the real world, functional data is usually recorded in finite, discrete points, it still has much higher dimension than traditional multivariate data. This unique feature makes extensions of existing multivariate methods to the context of functional data a challenging problem. Some of the proposed multivariate techniques have reasonable computational requirements in lower dimensions but will be computationally intractable in higher dimensional cases due to the curse of dimensionality.

One fundamental task in functional data analysis is to find the location, or in other word, the most representative center for a functional curve sample. This can potentially be solved by assigning a reasonable order for each curve within a sample. A natural way of ordering such curves is to measure the degree of centrality of each curve with respect to the underlying population distribution or given samples. This is the idea of statistical depth, whose aim is to provide a center-outward ranking within a data sample. 
The statistical depth was motivated by the idea of median in 1-D, and it was originally proposed in the multivariate framework to provide a center-outward ordering for multivariate data. The observation with a higher depth tends to be closer to the population center, while the datum with a lower depth is supposed to be far away from the center and less representative of the sample, which is possibly an outlier or abnormality.
 The well-known multivariate depths include the Tukeys halfspace depth~\citep{tukey1975mathematics}, the simplicial depth ~\citep{oja1983descriptive}, and the spatial depth ~\citep{chaudhuri1996geometric}.
 Some multivariate depths have been extended to functional data setting, such as the band depth  ~\citep{lopez2009concept}, the half-region depth ~\citep{lopez2011half}, and the functional spatial depth ~\citep{chakraborty2014data}. ~\citet{zuo2000general} studied the definitions, and properties, as well as categorized depths by their common properties. These depths can be utilized into multiple tasks, such as exploratory data analysis, robust location estimation, detecting outliers or building confidence intervals. Depths do not make assumptions about the inner structure of the data. They incorporates some intuitive ideas such as simplex, spatial signs, and functional bands to characterize the centrality. Besides, most depths have the merit of a low level of computation load.

\begin{figure}[h]
\begin{center}
\includegraphics[scale=0.6]{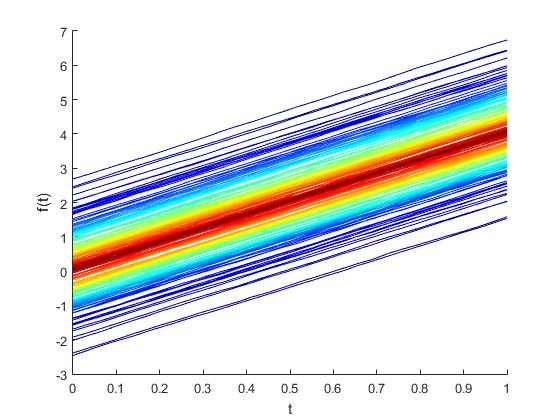}
\caption{A functional dataset colored by depth value}
\label{color_depth}
\end{center}
\end{figure}

As shown in figure \ref{color_depth}, depth value suggests how close the functional curves are away from the center, higher depth value means the curve is close to the distribution center and is intuitively 'deep' (colored red), while lower depth value implies that the curve is less representative of the distribution and far way from the center(colored blue). Thus, a depth-based trimmed mean where lower depth data will be excluded from the mean could be employed to estimate the location for the underlying distribution.

In this work, first some well-respected functional depths are introduced. Then a trimmed mean based on functional depth is constructed as a robust estimator of location for a functional data sample. Finally, Monte Carlo tests are conducted to study the performances of these estimators based on different functional depths.
\section{Method}
\subsection{State-of-the-Art depths for functional data}
\subsubsection{Half Region Depth}
Half region depth \citep{lopez2011half} is based on the the notion of 'half-region' which is determined by a functional curve. The two half regions determined by curve $x$ are illustrated in figure \ref{Illustration of half region} in a parallel coordinate. A functional curve $x$ divides the functional space into two complementary half regions: upper half region and lower half region. Half region depth $HRD(x;X) $ is equal to the smaller of the two probability mass carried by these two complementary half regions under distribution $X$. If the functional curve $x$ locates far away from the center of distribution $X$, then half region depth $HRD(x;X)$ will  be close to 0 since the smaller probability mass is close to 0. On the other hand, if $x$ locates close to the center of $X$, then half region depth $HRD(x;X)$ will be close to 0.5 since both probability masses over the complementary region are about 0.5. It use \emph{tailedness} to measure the curve centrality. Half region depth has computational advantages compare to other concepts of functional depth. It is mathematically defined as:
\begin{figure}[h]
\begin{subfigure}[h]{.5\textwidth}
  \centering
  \includegraphics[scale=0.3]{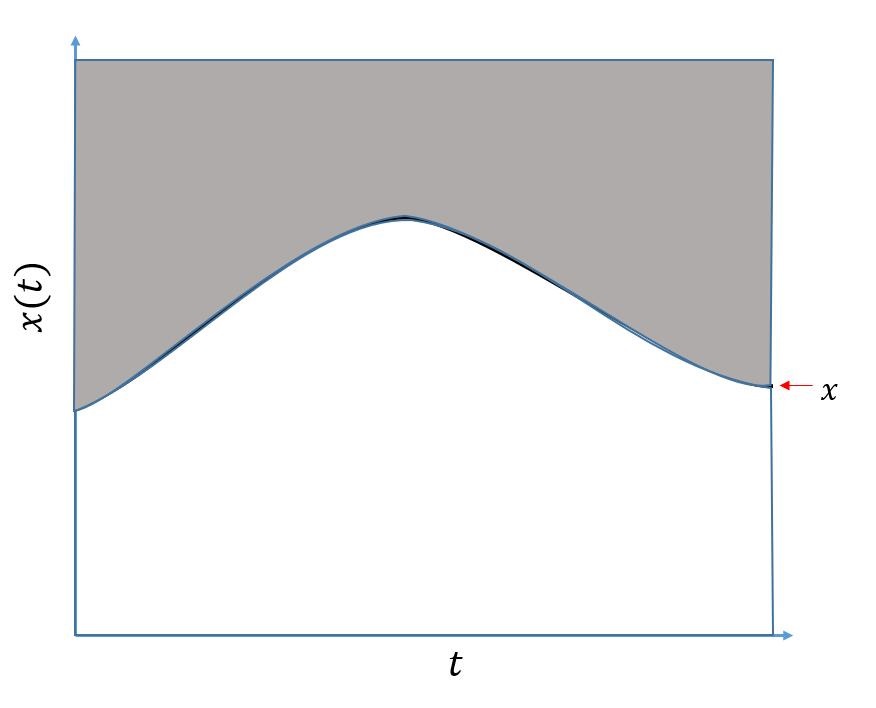}
  \caption{Upper half region of x}  
\end{subfigure}
\begin{subfigure}[h]{.5\textwidth}
  \centering
  \includegraphics[scale=0.3]{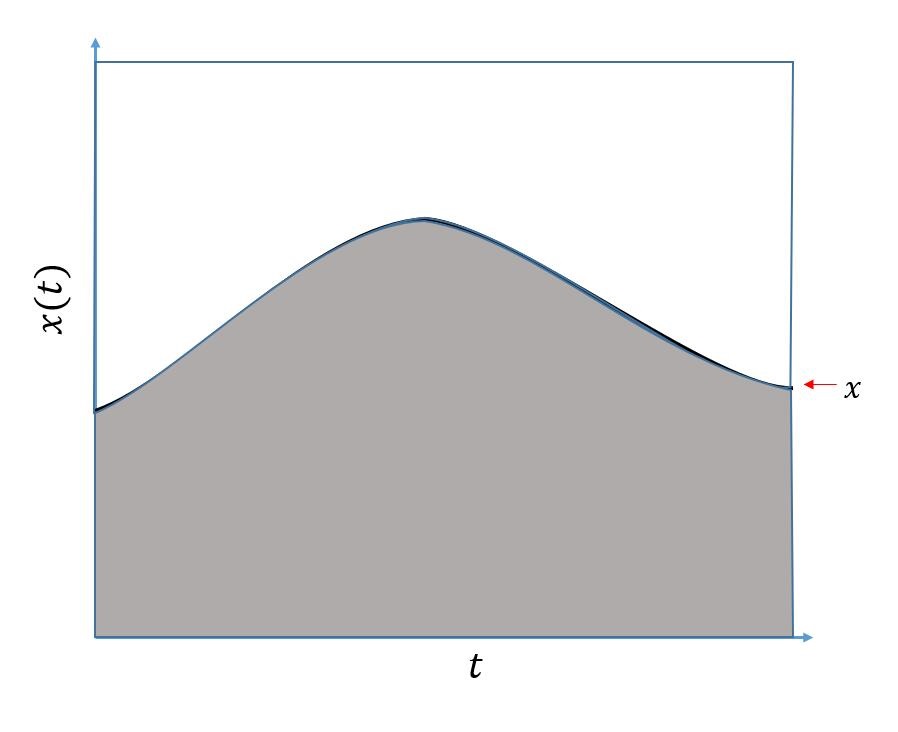}
  \caption{Lower half region of x}  
\end{subfigure}
\caption{Illustration of half region}
\label{Illustration of half region}
\end{figure}
\begin{definition}
Provided a target curve $x$ and reference distribution $X$, the half region depth $HRD(x;X)$ is defined as:
\[HRD(x;X)=\min(\mathbb{P}(X_r \geq x),\mathbb{P}(X_r \leq x))\]
\end{definition}
Here,$X_r$ is a random variable generated from distribution $X$. Compared with other functional depth, the half-region depth can be applied to high-dimensional data with little
computational cost.
\subsubsection{Functional Majority Depth}
Functional majority depth is extended from majority depth \citep{singh1991notion}. Assume $x_r\leftarrow X$ and it divides the functional space where $X$ lives into two regions, the majority half region of $x_r$ is the one of the two half regions that carries bigger probability mass. Functional majority depth measures the proportion of times that target curve $x$ is in the majority space given a reference distribution $X$.
\begin{definition}
Provided a target curve $x$ and a reference distribution $X$, functional majority depth $FMJ(x;X)$ is defined as:
\[FMJ(x;X)=\mathbb{P}(x \in MJS(X_r))\]
where $X_r$ is a random variable from $X$, and $MJS(X_r)$ denotes the majority half region of $X_r$.
\end{definition}
Functional majority depth shares some common features with half region depth. They both employ one curve to divide the functional space to two complementary half regions. While half region depth uses the tailedness (the half region carrying smaller probability mass) to describe the centrality, functional majority depth adopts the frequency of $x$ being in a majority half region (the half region carrying bigger probability mass)to describe the centrality. 
\subsubsection{Band Depth}
Band depth \citep{lopez2009concept} is based on the idea of functional band. In the parallel coordinate, the functional band is the area enclosed by $J$ curves. Its mathematical definition is:
\begin{figure}[h]
\begin{center}
\includegraphics[scale=0.4]{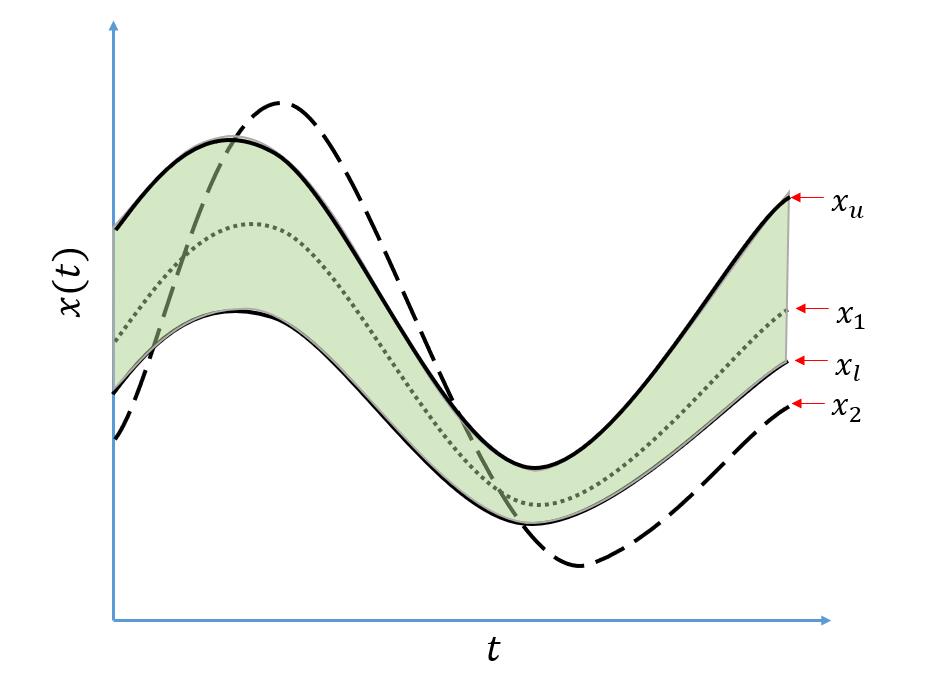}
\end{center}
\caption{Illustration of functional band}
\label{band illustration}
\end{figure}

\begin{definition}
\label{band definition}
In the parallel coordinate,$x_1(t)$,$x_2(t)$,...,$x_j(t)$ are $j$ functional curves defined over $t\in I$,the functional band $B(x_1(t),x_2(t),...,x_j(t))$ delimited by these $j$ curves is:
        	\[B(x_1,x_2,...,x_j)= \{t,x(t) :t\in I,\min_{r=1,2,...,j}x_r(t) \leq x(t)\leq \max_{r=1,2,...,j}x_r(t)\}\]
\end{definition}
A functional band delimited by 2 solid line curves $x_u,x_l$ is colored green as illustrated in figure \ref{band illustration}. The dotted line $x_1$ which is \emph{completely} inside the band and more similar to $x_u,x_l$ than the dashed curve $x_2$ which is \emph{partially} inside the band and has less similarity . Thus band can be used to measure the similarity between a functional curve and the boundaries of the band. Given a functional distribution $X$, band depth $ BD(x;X)$  can be viewed as the probability of the cases that  $x$  is \emph{completely} inside the band delimited by $j$ curves,  which are repeatedly randomly sampled from $X$.  The more bands that the target curve is completely inside, the larger the depth of the target curve would be. Band depth is defined as:
\begin{definition}
 If functional curve $x$ and functional distribution $X$ are both defined over continuum $I$, $X_1$,$X_1$,$...$,$X_j$ are $j$ independent random variables from the functional distribution $X$, the band depth $BD(x,X)$ is:
        \[BD^{(j)}(x,X)=Prob(x\in B(X_1,X_2,...,X_j))=\mathbb{E}(\mathds{1}(x\in B(X_1,X_2,...,X_j)))\]
        \[BD(x,X)=\sum_{j=2}^{J}BD^{(j)}(x,X)\]
\end{definition}
$J$ is chosen to be 3 by \citet{lopez2009concept}.
\subsubsection{Modified Band Depth}
Adopting the Lebesgue measure on $I$, band depth has been extended to a more flexible version: modified band depth \citep{lopez2009concept}. Modified band depth considers not only the cases when target curve is \emph{completely} in band but also the cases when target curve is \emph{partially} in band. First, Let:
\[
A_j\Big(x(t)\in B(x_1,x_2,...,x_j)\Big)=\{ t\in I:\min_{r=1,2,...,j}x_r(t)<x(t)< \max_{r=1,2,...,j}x_r(t) \}
\]
If $\lambda$ is the Lebesgue measure on $I$, let:
\[
\lambda_{r}\bigg(A_j\Big(x(t)\in B(x_1,x_2,...,x_j)\Big) \bigg)=\frac{\lambda\bigg(A_j\Big(x(t)\in B(x_1,x_2,...,x_j)\Big) \bigg)}{\lambda(I)}
\]
\begin{definition}
\label{modified_band_depth_definition}
 If functional curve $x$ and functional distribution $X$ are both defined over continuum $I$,$X_1,...,X_j$ are $j$ independent random variables from the functional distribution $X$, the modified band depth $MBD(x;X)$ is:
        \[ MBD(x,X)= \mathbb{E}\Bigg(\lambda_{r}\bigg(A_j\Big(x\in B(X_1,...,X_j)\Big) \bigg) \Bigg) \]
        $j$ is chosen to be 2 by \citet{lopez2009concept}.
\end{definition}
\subsubsection{Functional Spatial Depth}
\begin{figure}
\begin{center}
\includegraphics[scale=0.45]{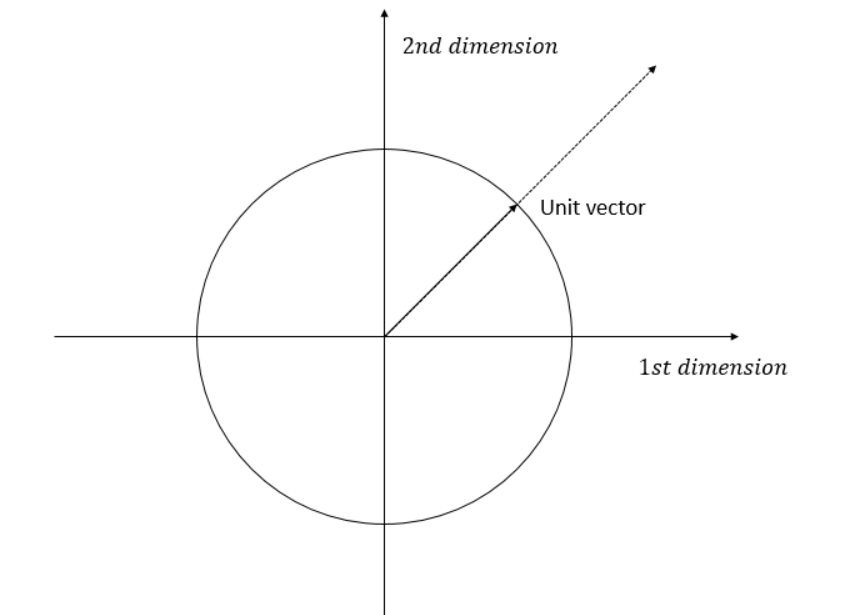}
\end{center}\
\caption{Illustration of spatial sign function in a 2-D space}
\label{unit vector}
\end{figure}

Functional spatial depth \citep{chakraborty2014data} is defined based on spatial sign function. The spatial sign function of $x \in \mathbb{R}^d$ is given by:
\[FS(x) = \left\{
        \begin{array}{ll}
            \dfrac{x}{\|x\|} & \quad x \neq 0 \\
            0 & \quad x = 0
        \end{array}
    \right.\]    
Spatial sign function will reduce a vector to a unit vector as shown in figure \ref{unit vector}. This process remove the distances from the spatial depth.
\begin{definition}
Provided a target curve $x$ and a reference distribution $X$, functional spatial depth $FSD(x;X)$ is defined as:
\[
FSD(x;X)=1- \|\mathbb{E}(FS(x-Y))\|
\]
Here,$Y$ is a functional random variable with probability distribution $X$.
\end{definition}
\subsubsection{Remarks on functional depth: Global or Local?}
Our comparison is restricted into ``global depth". Under global depth, the ``center" consists of a set of curves globally maximizing the depth values. Global depth tends to ignore multimodality features of the underlying distribution
$P$. On the other hand, some functional depth appears to be sensitive to multimodality features and local maxima. In these depths, some properties are compromised such as center-outward order property,inner curve can have a lower depth.

\begin{figure}[h]
 \begin{center}
 \includegraphics[scale=0.5]{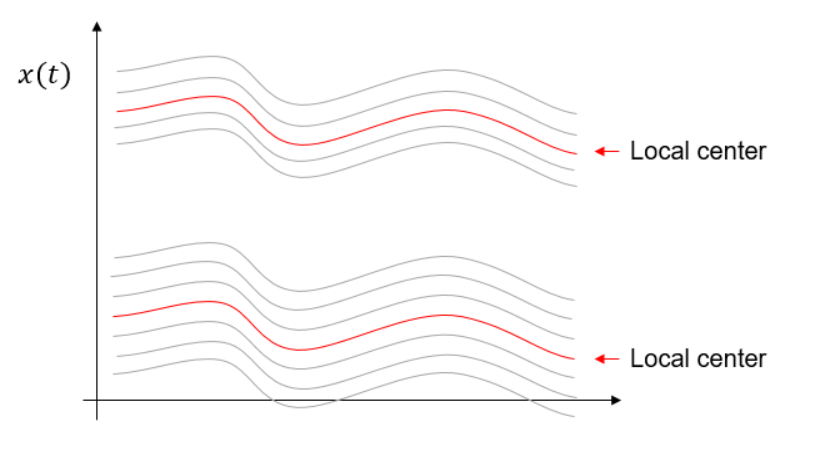}
 \end{center}
 \caption{$h$-mode depth in a functional distribution with 2 modals }
 \label{h-mode multimodal dist}
 \end{figure}

\begin{figure}
\centering
\begin{subfigure}{.5\textwidth}
  \centering
  \includegraphics[width=6cm,height=8cm ]{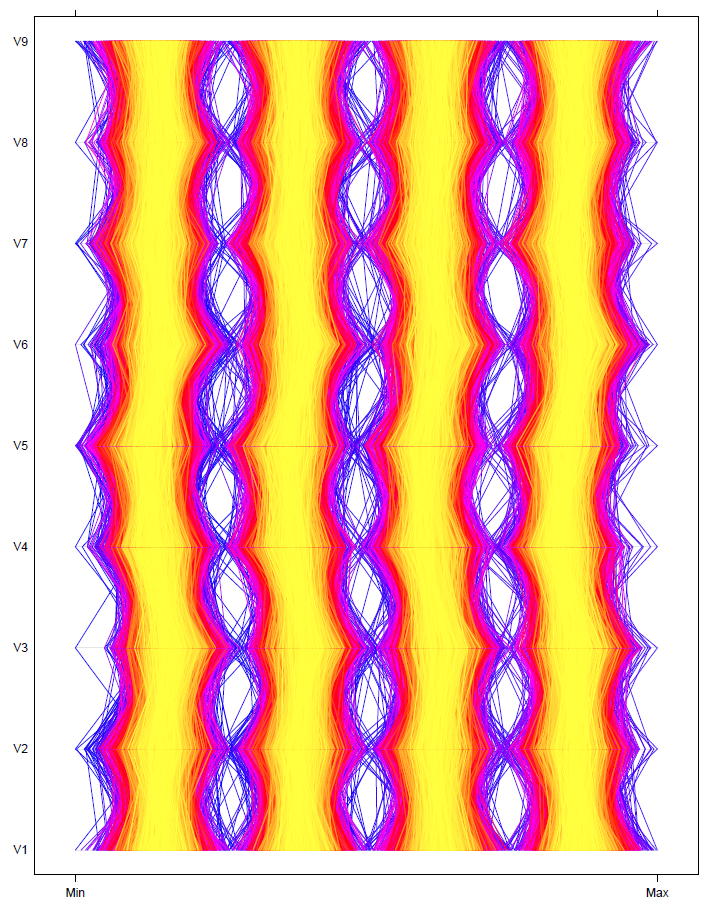}
  \caption{Local depth}
  \label{fig:sub1}
\end{subfigure}%
\begin{subfigure}{.5\textwidth}
  \centering
  \includegraphics[width=6cm,height=8cm ]{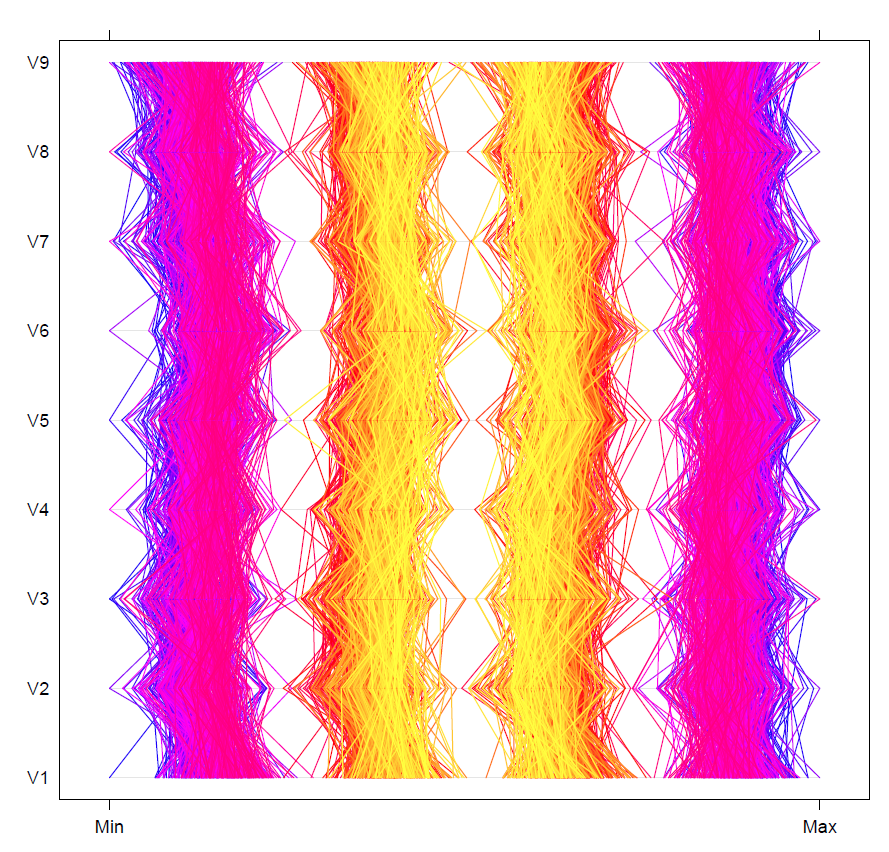}
  \caption{Global depth}
  \label{fig:sub2}
\end{subfigure}
\caption{Local depth and global depth identifying the center}
\label{local vs global}
\end{figure}

$h$-mode depth  \citep{cuevas2007robust} is such an example. It is a distance-based functional depth. The depth value decreases as the distance between the target curve and reference population increases. $h$-mode depth is sensitive to the multimodality features of the distribution. In figure \ref{h-mode multimodal dist}, the center curve has a lower depth and the highest depth occurs at the local centers marked with red curves.\\
\begin{definition}
Provided a target curve $x$ and a reference distribution $X$, $h$-mode depth $HMD(x;X)$
 \[HMD(x;X)=\mathbb{E}(\frac{exp(\|x-X_r\|)}{h})\]
 Here, $X_r$ is a random curve from distribution $X$.
\end{definition}
It is noteworthy that $h$-mode depth is not a depth in a strict way. Generally, depth is motivated from a non-parametric perspective where distances are excluded. Take a look at spatial depth, the distances along each direction are normalized out and only a unit vector which denotes a direction remains in the final depth. Some desired properties, such as maximum at symmetric distribution center, came along with this motivation and gradually became what researcher would expect a depth should have. Detail discussions of these properties are out the scope of this work. For $h$-mode depth, due to the incorporation of distance into the depth construction, $h$-model depth doesn't carry some of the desired  properties of depth. Figure \ref{local vs global} illustrates the difference when local depth and global depth are applied to the same functional dataset. The heatmap color represents the depth value. Local depth identifies the 4 local centers in the left figure, but it also assigns lower depth values to curves in the global center. On the other hand, though global depth ignores the local center, it succeeds in detecting the global center.

\section{Simulation}
\subsection{Location estimator based on functional depth}
Functional depth provides a center-outward order for a sample of curves. Given a functional distribution , the curve with a lower depth usually tends to be far away from the distribution center and could be considered less typical and more abnormal. The curves with higher depth are more representative of the distribution compared with curves with lower depth. Thus, by trimming out the curves with least $\alpha$  percent depth in a sample, a $\alpha$-trimmed mean based on depth is constructed to estimate the distribution location. Given a functional curves sample containing $n$ curves: $x_1,x_2,x_3,...,x_n$, the robust location estimator $\hat{m}_n^\alpha$ based on depth is:
\[ \hat{m}_n^\alpha = \dfrac{ \sum \displaylimits_{i=1}^{n-[n\alpha]}x_{(i)}}{n-[n\alpha]}\]
where $x_{(i)}$ is the sample ordered from the deepest to the least deep curve and $[n\alpha]$ is the integer part of $n\alpha$. We tested the performance of robust estimators based on 5 functional depth methods: half region depth, functional majority depth, band depth, modified band depth and functional spatial depth. We also computed the performance of the untrimmed mean as a  comparison baseline for other robust estimators. 
\subsection{Models}
In this subsection, testing models are introduced. Each model is a mixture model of normal curves and outlier curves in $\mathbb{H}$:
 \[ Y_{mix} = \begin{cases} 
          Y_{nor} &\mathrm{with\  probability\ }  1-q \\
          Y_{out} &\mathrm{with\  probability\ }  q
       \end{cases}
    \]
Here, $q$ is the contamination probability(usually small). There is $S=1000$ replications for each model. Each model in each replication will generate $n=50$ curves, which is expected to compose with $nq$ outlier curves and $n(1-q)$ normal curves. Depth value for each curve, calculated with respect to the sample, is used to measure its relative centrality. Curves with low depths will be trimmed out, and depth-based trimmed mean will be used as location estimator for the sample.

\begin{figure}
\begin{center}
\includegraphics[scale=.45]{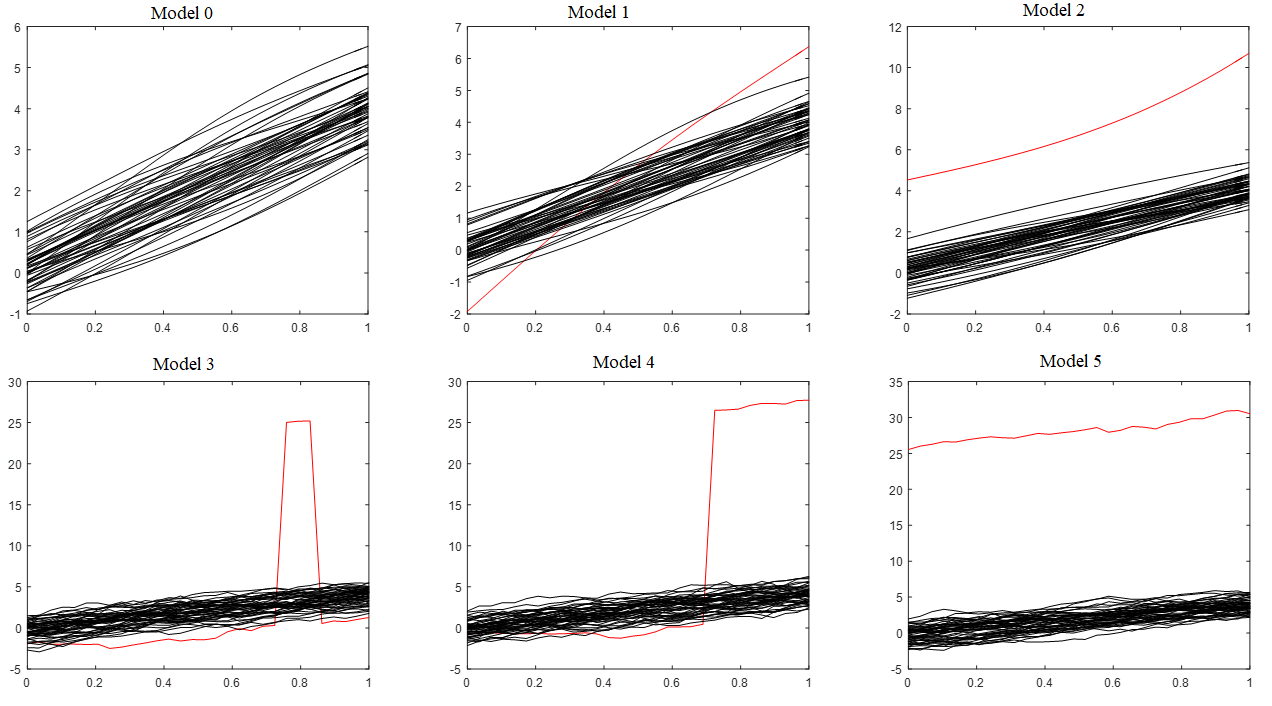}
\end{center}
\caption{Models plot for robust estimator simulation test, each figure includes 49 normal curves(black) and 1 outlier curve(red)}
\label{robust test models}
\end{figure}

Below are 6 test models, as illustrated in figure \ref{robust test models}. The detail specifications of these models are as below,

\begin{itemize}

\item Model 0.\\
Model 0 is defined as: 
$$X_i(t)=g(t)+e_i(t),1 \leq i \leq n $$
Where the mean $g(t)=4t$, $ t $ is a vector corresponding 30 equally spaced points in interval $[0,1]$ . $e_i(t)$ represents the random noise, and it is a stochastic Gaussian process with zero mean and covariance function $\gamma(s,t)=e^{-0.45|t-s|}$. Model 0 has no outlier curves. Normal curves of models 1 and 2 will be generated from MM0.\citep{sguera2014spatial}  The normal curves for model 4,5 and 6 has the same mean as model 0 but a relatively higher noise and appear to be less smooth \citep{lopez2009concept}.
\item Model 1.\\
Model 1 and 2 shares the same normal curves which are from model 0. 


Model 1 contaminated by outliers with different trend:$$Y_i(t)=(1-c_i)X_i(t)+c_i(8t-2+e_i(t))$$ where $c_i$ is 1 with probability $q$ and 0 with probability $1-q$. 
\item Model 2.\\
Model 2 contaminated by outliers with different shape and location compared with normal curves: $$Y_i(t)=(1-c_i)X_i(t)+c_i(4e^t+e_i(t))$$
\item Model 3.\\
The normal curves of model 3,4 and 5 came from same stochastic process.   \citep{lopez2009concept}
\[ X_i'(t)=g(t)+e_i'(t) \]
where  $e_i'(t)$ is a zero-mean Gaussian component with covariance function:$\gamma(s,t)=e^{-(s-t)^2}$.
Model 3 is contaminated by peaks: $Y_i(t)=X_i'(t)+c_i\sigma_iK$,if $T_i \leq t \leq T_i +l,1\leq i \leq n$ and $Y_i(t)=X_i'(t)$ if $t \not \in [T_i, T_i+l]$, where $l=2/30$ and $T_i$ is a random number from a uniform distribution in $[0,1-l]$. $K=25$ is a contamination size constant.$c_i$ is 1 with probability $q$ and 0 with probability $1-q$. $\sigma_i$ is a sequence of random variables independent of $c_i$ taking values 1 and -1 with probability 0.5.
\item Model 4.\\
Model 4 is partially contaminated: $Y_i(t)=X_i'(t)+c_i\sigma_iK$,if $t\geq T_i, 1\leq i \leq n$, and $Y_i(t)=X_i'(t)$ if $t<T_i$, where $T_i$ is a random number generated from a uniform distribution on $[0,1]$. $ K, c_i $ and $ \sigma_i $ are the same as in model 3.
\item Model 5.\\
Model 5 includes a parallel shift contamination:$Y_i(t)=X_i'(t)+c_i\sigma_iK$, where $c_i$ is 1 with probability $q$ and 0 with probability $1-q$, $ K, c_i $ and $ \sigma_i $ are the same as in model 3.
\end{itemize}
\subsection{Evaluation Criterion}
For each model, we considered $S=1000$ replications.In each replication, we generated $n=50$ curves. Each curve is recorded in $T=30$ equidistant points in $[0,1]$.  The contamination probability is $q =0.1$. We chose the trimmed percentage $\alpha=20\%$. Integrated square error is utilized \citep{lopez2009concept} to evaluate the performance. The integrated squared error for each replication is defined as:
\[ISE(j)=\frac{1}{T} \sum \displaylimits_{k=1}^{T}\big[ \hat{g}_n(k/T)-g(k/T)\big]^2\]
Here, $\hat{g}_n(k/T)$ is the corresponding location estimator.$j$ means the $j^{th}$ replication. $T$ denotes the number of dimensions of functional curve in the test models.
\subsection{Result}
\begin{table}[htbp]
  \centering
  \caption{Mean integrated squared errors using robust location estimators based on functional depth}\label{robust estimation result}
    \begin{tabular}{c c c c c c c}
    \hline \hline
    \textbf{Method} & \textbf{Model 0} & \textbf{Model 1} & \textbf{Model 2} & \textbf{ Model 3} & \textbf{Model 4} & \textbf{Model 5} \\
    \hline
    \textbf{Half region depth} & 0.0293 & 0.0119 & 0.1426 & 0.1412 & 0.6145 & 0.4172 \\
    \hline
    \textbf{Functional majority depth} & 0.0327 & 0.0150 & 0.0761 & 0.1806 & 0.2753 & 0.1954 \\
    \hline
    \textbf{Band depth} & 0.0269 & 0.0096 & 0.1719 & 0.0971 & 0.2865 & 0.6717 \\
    \hline
    \textbf{Modified band depth} & 0.0255 & 0.0100 & 0.0402 & 0.1757 & 0.2159 & 0.0445 \\
    \hline
    \textbf{Functional spatial depth} & 0.0240 & 0.0085 & 0.0500 & 0.0256 & 0.0286 & 0.0551 \\
    \hline
    \textbf{Untrimmed Mean} & 0.0201 & 0.0208 & 0.2902 & 0.1455 & 0.6572 & 1.2619 \\
    \hline
    \end{tabular}%
  \label{tab:addlabel}%
\end{table}%
\paragraph{ } 
Table \ref{robust estimation result} is the simulation result. In model 0, untrimmed mean is the best location estimator since model 0 is not contaminated with outliers. Untrimmed mean as a location estimator preserves the most information when building the location estimator. Estimator based on functional spatial depth performance ranks the best in model 1,3 and 4, and ranks 2nd in model 2 and 5. Estimator based on modified band depth also works well for the test models. Its performance ranks the best in model 2,5. The performance of estimator based on half region depth and functional majority depth are relatively worse, but they are still better than untrimmed mean in the contaminated models except for model 3. Functional spatial depth delivers a comparable performance in model 2.

In model 1 through 5, the outlier has a different shape or location than normal curves. When the test curve is the outlier and it appears away from the center, then the absolute value of sum of spatial sign function will accumulate and increase significantly, resulting into a clear drop in depth value for test curve.
As for modified band depth, its sensitivity to outlier curve is due to its close relationship with multivariate simplicial depth \citep{liu1990notion}. Roughly speaking, the simplicial depth measures the probability of cases where target point $ x \in \mathbb{R}^d $ is inside the multivariate simplices which is formed by points from reference distribution in $ \mathbb{R}^d $. Modified band depth can be viewed as average of 1-D simplicial depth over all dimensions. When the test is curve shifted from the center, each 1-D simplicial depth will decrease leading to a lower modified band depth.

\begin{figure}[h]
        \begin{center}
        \includegraphics[scale=0.5]{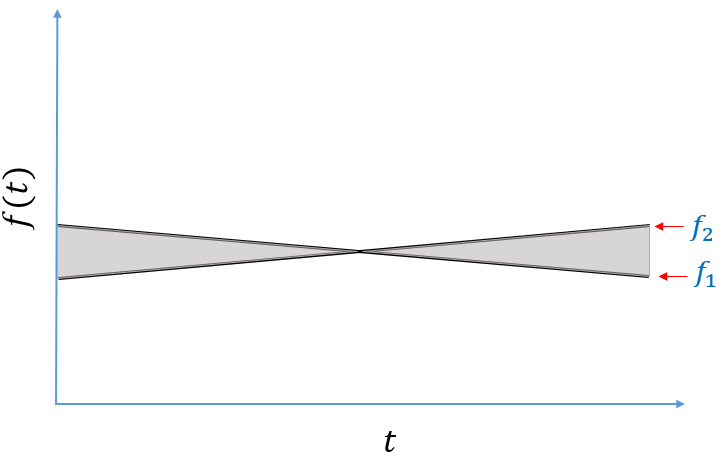}
        \end{center}
        \caption{ Functional band when 2 boundaries are crossing}
        \label{crossing_band}
        \end{figure}

Although band depth is claimed to be convenient in detecting curve shape abnormality\citep{liu1990notion} , the performance is not outstanding compared with other methods. This is because band depth only counts the case when target curve is completely inside the band, while it rules out cases when target curve is partially in band. In figure \ref{crossing_band},the functional band is delimited by two curve boundaries which are \emph{crossover} with each other. The probability that a test functional curve is completely inside the band $B(f_1,f_2)$ is 0. In the tested models, the noise included in normal curves leads to crossover band frequently, which band depth doesn't count. This hurts the performance of band depth, and  it sometimes fails to distinguish abnormal shape curve under the tested models.

Regarding functional majority depth, it performs badly in model 3. This is again due to the\emph{ crossover} between curves. A test curve will have lower chances of being \emph{completely} inside the majority region ,which leads to a lot of 0 depth. Eventually, a lot of normal curves will have lower depths and this lowers the performance of functional majority depth. However, in model 5, functional majority depth achieves a comparable performance since the outlier curve is significantly shift from the center and its depth is always 0 by visual inspection. The performance is even better in model 2, where normal curve noise is smaller and \emph{crossovers} between normal curves are less.

\section{Conclusions}
\paragraph{ } 
Depth is a function that maps datum in $\mathbb{R}^d$ to a non-negative real number. It provides a distribution-based central-outward order for the data. Depth has been extended to functional data, and it provides an intuitive, relative fast way to characterize the most representative center curve as well as detecting some possible abnormalities.
By trimming out functional curves with least depth, robust location estimators based on several functional depth have been built and their performance are compared. 

Based on the test results, estimators  based on functional spatial depth and modified  band depth performed well consistently in most tested models. While functional spatial depth is sensitive to shape outlyingness due to spatial function, modified band depth sensitivity comes from 1-D simplicial depth.
Estimator based on band depth also works better than mean and shows certain sensitivity to curve shape outlyingness .Estimators based on half region depth and functional majority depth show the relatively worse performances in the 5 compared depth-based estimators, but they still surpass the estimator based on untrimmed mean for all contaminated models except model 3. This is because band depth, half region depth and functional majority depth are vulnerable to \emph{crossovers} between curves and will assign unexpected low depth values to normal curves. 
\clearpage
\bibliographystyle{apalike}

\bibliography{wmbd_ref}

\begin{thebibliography}{11}
\providecommand{\natexlab}[1]{#1}
\providecommand{\url}[1]{\texttt{#1}}
\expandafter\ifx\csname urlstyle\endcsname\relax
  \providecommand{\doi}[1]{doi: #1}\else
  \providecommand{\doi}{doi: \begingroup \urlstyle{rm}\Url}\fi

\bibitem[Chakraborty and Chaudhuri(2014)]{chakraborty2014data}
Anirvan Chakraborty and Probal Chaudhuri.
\newblock On data depth in infinite dimensional spaces.
\newblock \emph{Annals of the Institute of Statistical Mathematics},
  66\penalty0 (2):\penalty0 303--324, 2014.

\bibitem[Chaudhuri(1996)]{chaudhuri1996geometric}
Probal Chaudhuri.
\newblock On a geometric notion of quantiles for multivariate data.
\newblock \emph{Journal of the American Statistical Association}, 91\penalty0
  (434):\penalty0 862--872, 1996.

\bibitem[Cuevas et~al.(2007)Cuevas, Febrero, and Fraiman]{cuevas2007robust}
Antonio Cuevas, Manuel Febrero, and Ricardo Fraiman.
\newblock Robust estimation and classification for functional data via
  projection-based depth notions.
\newblock \emph{Computational Statistics}, 22\penalty0 (3):\penalty0 481--496,
  2007.

\bibitem[Liu(1990)]{liu1990notion}
Regina~Y Liu.
\newblock On a notion of data depth based on random simplices.
\newblock \emph{The Annals of Statistics}, pages 405--414, 1990.

\bibitem[L{\'o}pez-Pintado and Romo(2009)]{lopez2009concept}
Sara L{\'o}pez-Pintado and Juan Romo.
\newblock On the concept of depth for functional data.
\newblock \emph{Journal of the American Statistical Association}, 104\penalty0
  (486):\penalty0 718--734, 2009.

\bibitem[L{\'o}pez-Pintado and Romo(2011)]{lopez2011half}
Sara L{\'o}pez-Pintado and Juan Romo.
\newblock A half-region depth for functional data.
\newblock \emph{Computational Statistics \& Data Analysis}, 55\penalty0
  (4):\penalty0 1679--1695, 2011.

\bibitem[Oja(1983)]{oja1983descriptive}
Hannu Oja.
\newblock Descriptive statistics for multivariate distributions.
\newblock \emph{Statistics \& Probability Letters}, 1\penalty0 (6):\penalty0
  327--332, 1983.

\bibitem[Sguera(2014)]{sguera2014spatial}
Carlo Sguera.
\newblock Spatial depth-based methods for functional data.
\newblock 2014.

\bibitem[Singh(1991)]{singh1991notion}
K~Singh.
\newblock A notion of majority depth.
\newblock \emph{Unpublished document}, 1991.

\bibitem[Tukey(1975)]{tukey1975mathematics}
John~W Tukey.
\newblock Mathematics and the picturing of data.
\newblock In \emph{Proceedings of the International Congress of Mathematicians,
  Vancouver, 1975}, volume~2, pages 523--531, 1975.

\bibitem[Zuo and Serfling(2000)]{zuo2000general}
Yijun Zuo and Robert Serfling.
\newblock General notions of statistical depth function.
\newblock \emph{Annals of statistics}, pages 461--482, 2000.

\end{thebibliography}
\end{document}